\documentclass{aa}
\usepackage{txfonts}
\usepackage{graphicx}
\begin{document}

\title{The radii of the nearby K5V and K7V stars 61\,Cyg\,A\,\&\,B}
\subtitle{CHARA/FLUOR interferometry and CESAM2k modeling}
\titlerunning{Interferometric observations of 61 Cyg A \& B with CHARA/FLUOR}
\authorrunning{P. Kervella et al.}
\author{
P.~Kervella\inst{1} \and A.~M\'erand\inst{2} \and B. Pichon\inst{3} \and 
F.~Th\'evenin\inst{3} \and U.~Heiter\inst{4} \and L.~Bigot\inst{3} \and T.~A.~ten 
Brummelaar\inst{2} \and H.~A.~McAlister\inst{2} \and S.~T.~Ridgway\inst{5} \and 
N.~Turner\inst{2} \and J.~Sturmann\inst{2} \and L.~Sturmann\inst{2} \and 
P.~J.~Goldfinger\inst{2} \and C.~Farrington\inst{2}
}
\offprints{P. Kervella}
\mail{Pierre.Kervella@obspm.fr}
\institute{LESIA, Observatoire de Paris, CNRS\,UMR\,8109, UPMC, Universit\'e Paris Diderot, 5 Place Jules Janssen, 92195 Meudon, France
\and
Center for High Angular Resolution Astronomy, Georgia State University, PO Box 3965, Atlanta, Georgia 30302-3965, USA
\and
Universit\'e de Nice-Sophia Antipolis, Lab. Cassiop\'ee, UMR\,6202, Observatoire de la C\^ote d'Azur, BP 4229, 06304 Nice, France
\and 
Department of Physics and Astronomy, Uppsala University, Box 515, SE-751 20  Uppsala, Sweden
\and
National Optical Astronomy Observatories, 950 North Cherry Avenue, Tucson, AZ 85719, USA
}
\date{Received ; Accepted}
\abstract
{The main sequence binary star 61\,Cyg (K5V+K7V) is our nearest stellar neighbour in the northern hemisphere.
This proximity makes it a particularly well suited system for very high accuracy interferometric radius measurements.}
{Our goal is to constrain the poorly known evolutionary status and age of this bright binary star.}
{We obtained high accuracy interferometric observations in the infrared K$^\prime$ band, 
using the CHARA/FLUOR instrument.  We then computed evolutionary models of 61\,Cyg~A \& B
with the CESAM2k code. As model constraints, we used a combination of observational parameters from
classical observation methods (photometry, spectroscopy) as well as our new interferometric radii.}
{The measured limb darkened disk angular diameters are $\theta_{\rm LD}(A) = 1.775 \pm 0.013$\,mas 
and $\theta_{\rm LD}(B) = 1.581 \pm 0.022$\,mas, respectively for 61\,Cyg~A and B. Considering the
high accuracy parallaxes available, these values translate into photospheric radii of
${\rm R}(A) = 0.665 \pm 0.005\ {\rm R}_{\odot}$ and ${\rm R}(B) = 0.595 \pm 0.008\ {\rm R}_{\odot}$.
The new radii constrain efficiently the physical parameters adopted for the modeling of both stars, allowing
us to predict asteroseismic frequencies based on our best-fit models.}
{The CESAM2k evolutionary models indicate an age around 6\,Gyrs and are compatible with
small values of the mixing length parameter. The measurement of asteroseismic oscillation frequencies
in 61\,Cyg~A \& B would be of great value to improve the modeling of this important fiducial stellar system,
in particular to better constrain the masses.}

\maketitle

\section{Introduction}

Binary stars are numerous in the Galaxy but only few of them can be fully calibrated physically. 
They are interesting for several reasons, among them the most important for evolutionary 
modeling is that the two components are coeval, and share the same chemical abundances
of the original material. This makes their modeling 
much easier than for single stars, 
as several free parameters of the models are shared between the two stars (age and helium 
content). Possibly the best example of physical calibration of 
a binary system is provided by our closest neighbour, $\alpha$\,Cen (Th\'evenin et 
al.~\cite{thevenin02}). Its two solar-type components were analyzed in detail with the use of photometry, 
spectrometry, astrometry and interferometry. 
It resulted in excellent knowledge of the masses of the two components (within less than 1\%), 
and, through modeling of the stars, of fundamental parameters not directly accessible to
measurements, like the age or the helium  content. 
Therefore, these two stars are, with Procyon, Sirius and Vega, among the most important 
benchmark stars for new developments of stellar structure and atmosphere models.
The present study focusses on 61\,Cyg~A (\object{HD 201091}, \object{HIP 104214})
and B (\object{HD 201092}, \object{HIP 104217}), the nearest
stars in the northern hemisphere. The two stars constitute a visual binary
pair with a very long orbital period ($\approx$ 650\,yrs),
also known as \object{Gl 820}.
The parallax of ths system was first measured by Bessel~(\cite{bessel1838}), and it is now known 
with extremely high accuracy. 
The proper motion of more than 5${\arcsec}$ per year, first determined by Piazzi in the XVIII$^{\rm th}$ century, 
makes it one of the fastest moving stars in terms of apparent displacement. Although some of this 
motion comes from the proximity of 61\,Cyg to us, the pair
has also a high radial velocity of 108 km/s, indicating that 61\,Cyg is not a member of the thin disk of our Galaxy.
The proximity of 61\,Cyg makes it a northern analog of the numerical modeling benchmark 
$\alpha$\,Cen. Its large parallax also means that this is the easiest low-mass dwarf to resolve
interferometrically. 
The spectral types of its two members (K5V and K7V) ideally complement our previous studies 
of $\alpha$\,Cen A \& B (G2V+K1V, Kervella et al.~\cite{kervella03a}, Bigot et al.~\cite{bigot06}).
The masses of 61\,Cyg A \& B are controversial, with estimates ranging from 
approximately 0.74 and 0.46 M$_\odot$ (Gorshanov et al.~\cite{gorshanov06}) to 0.67 and
0.59\,M$_\odot$ (Walker et al.~\cite{walker95}). With effective temperatures of about 4\,400 and 4\,000\,K,
they shine with luminosities of only 0.15 and 0.08\,L$_\odot$. 
There is no confirmed planet around them, although indications exist that 61\,Cyg~B could host a
giant planetary companion (see. Sect.~\ref{discussion}). 
As dimmer versions of solar type stars, they have magnetic cycles similar to that of the Sun, 
the brighter 8 years, the fainter 11 years (Hempelmann et al.~\cite{hempelmann06}).
Their rotation periods are of the order of 35 days, therefore no rotational distorsion of their photospheres
is expected. The abundances of heavy chemical elements have 
been determined (Luck \& Heiter~\cite{luck05}, \cite{luck06}) in these stars which are found slightly metal poor
(-0.2\,dex), so a priori older than the Sun but belonging to the galactic disk.

In Sect.~\ref{observations}, we detail the interferometric observations and 
the corresponding physical parameters we derive (angular diameters and linear radii).
Together with the additional observables listed in Sect.~\ref{constraints}, we propose in
Sect.~\ref{modelfit} a modeling of the two stars using the CESAM2k code. We finally
present asteroseismic frequency predictions in Sect.~\ref{astero}.

\section{Interferometric observations \label{observations}}

\subsection{Instrumental setup}

Our observations of 61\,Cyg were undertaken in November 2006 in the
near infrared $K'$ band ($1.9\leq \lambda \leq  2.3\;\mu\mathrm{m}$), at the CHARA Array 
(ten Brummelaar et al.~\cite{tenbrummelaar05}) using FLUOR, the Fiber Linked Unit for 
Optical Recombination (Coud\'e du Foresto  et al.~\cite{coude03}).
We used the FLUOR Data reduction software (DRS) (Coud\'e du Foresto et 
al.~\cite{coude97}; Kervella et al.~\cite{kervella04a}; M\'erand et
al.~\cite{merand06}), to extract the squared instrumental visibility of the interference fringes. 
The baseline was chosen according to the predicted angular sizes of 61\,Cyg\,A \& B 
(approximately 1.5 to 2 mas) and the wavelength of 
observation, in order to obtain the most constraining measurements on the angular diameters. 
This led to the choice of the CHARA telescopes S1 and E2, separated by a linear distance of 279\,m.
The calibrator stars were chosen in the catalogue compiled by M\'erand et al.~(\cite{merand05}), 
using criteria defined by these authors (Table~\ref{calibrators}). They were observed immediately before
or after 61\,Cyg in order to monitor the interferometric transfer function of the instrument.

We selected these calibrators so that their visibility measurements have comparable
signal-to-noise ratios to 61\,Cyg A \& B. As they have nearly the same effective
temperature and brightness as our targets, their angular diameters are also similar.
It results that the final precision of our angular diameter measurements of 61\,Cyg is limited
by the uncertainties on the calibrators' diameters.
We took them carefully into account in our final error bars, including the correlations between
different calibrated observations of the same star.
The reader is referred to Perrin~(\cite{perrin03}) for a detailed description of
the error propagation method we used. The final calibrated visibilites of 61\,Cyg\,A and B are
listed in Table~\ref{visib-table}.

As a remark, using smaller (hence fainter) calibrator stars would
have led to a larger observational uncertainty on the measured visibility of the calibrator and would
not have improved, or only on a marginal note, the accuracy of the angular diameters of 61\,Cyg A \& B.
\begin{table*}
      \begin{tabular}{llllccc}
        \hline \hline
        Star & $m_V$ & $m_{Ks}$ & Spect. & $\theta_{\rm LD}$ (mas) & $\theta_{\rm UD\,K}$ (mas) & $\alpha$\,($^\circ$) \\
        \hline
        \noalign{\smallskip}
        \object{HD 196753} & 5.94 & 3.34 & K0II-III & $1.047 \pm 0.013$ & $1.022 \pm 0.013$ & 16.2 \\
        \object{HD 200451} & 7.23 & 2.90 & K5III & $1.485 \pm 0.019$ & $1.441 \pm 0.019$ & 12.3 \\
        \hline
      \end{tabular}
    \caption{Calibrators used for the observations. They were selected in the catalogue
    assembled by M\'erand et al.~(\cite{merand05}).
    The limb darkened ($\theta_{\rm LD}$)
    and uniform disk ($\theta_{\rm UD\,K}$, for the $K$ band) angular diameters are
    given in milliarcseconds (mas). The angular separation $\alpha$ between the calibrators
    and 61\,Cyg is given in the last column, in degrees.
    }
    \label{calibrators}
\end{table*}


\begin{table}
\caption[]{Squared visibility measurements obtained for 61\,Cyg\,A \& B.
$B$ is the projected baseline length, and ``PA" is the azimuth of the
projected baseline (counted positively from North to East).
\label{visib-table}}
\begin{tabular}{ccrrc}
\hline \hline
MJD & Star & $B$ (m) & PA\,($^\circ$) & $V^2 \pm \sigma(V^2)$ \\
\hline
\noalign{\smallskip}
53540.352 & 61\,Cyg\,A & 210.98  & 5.35 & $0.1194 \pm 0.0092$ \\
53540.375 & 61\,Cyg\,A & 210.98  & 0.21 & $0.1381 \pm 0.0098$ \\
53540.429 & 61\,Cyg\,A & 210.89 & -11.70 & $0.1322 \pm 0.0082$ \\
53542.433 & 61\,Cyg\,A & 210.80 & -13.88 & $0.1356 \pm 0.0084$ \\
54043.158 & 61\,Cyg\,A & 205.50 & -32.51 & $0.1407 \pm 0.0258$ \\
54043.203 & 61\,Cyg\,A & 197.66 & -39.53 & $0.1882 \pm 0.0122$ \\
54044.149 & 61\,Cyg\,A & 206.30 & -31.35 & $0.1393 \pm 0.0270$ \\
54044.190 & 61\,Cyg\,A & 199.81 & -38.07 & $0.1792 \pm 0.0207$ \\
54052.124 & 61\,Cyg\,A & 33.54 & -21.29 & $0.9559 \pm 0.0123$ \\
54052.159 & 61\,Cyg\,A & 33.04 & -27.24 & $0.9239 \pm 0.0166$ \\
54052.193 & 61\,Cyg\,A & 32.23 & -32.48 & $0.9517 \pm 0.0092$ \\
54052.234 & 61\,Cyg\,A & 30.70 & -37.98 & $0.9656 \pm 0.0090$ \\
54055.169 & 61\,Cyg\,A & 96.60 & 72.69 & $0.6990 \pm 0.0107$ \\
54055.211  & 61\,Cyg\,A & 85.78 & 60.16 & $0.7427 \pm 0.0105$ \\
\hline
\noalign{\smallskip}
53540.401 & 61\,Cyg\,B & 210.98 &  -5.67 & $0.1714 \pm 0.0114$ \\
53540.449 & 61\,Cyg\,B & 210.67 & -16.13 & $0.1840 \pm 0.0116$ \\
54043.169 & 61\,Cyg\,B & 204.09 & -34.24 & $0.2982 \pm 0.0257$ \\
54044.159 & 61\,Cyg\,B & 205.08 & -33.05 & $0.3049 \pm 0.0162$ \\
54044.199 & 61\,Cyg\,B & 197.90 & -39.38 & $0.2694 \pm 0.0340$ \\
54049.172 & 61\,Cyg\,B & 231.56 &  52.54 & $0.1386 \pm 0.0092$ \\
54049.207 & 61\,Cyg\,B & 217.61 &  41.37 & $0.1919 \pm 0.0098$ \\
54052.135 & 61\,Cyg\,B & 33.41 &  -23.29 & $0.9670 \pm 0.0168$ \\
54052.171 & 61\,Cyg\,B & 32.80 &  -29.16 & $0.9405 \pm 0.0140$ \\
54052.208 & 61\,Cyg\,B & 31.76 &  -34.56 & $0.9253 \pm 0.0144$ \\
54055.180 & 61\,Cyg\,B & 94.08  &  69.75 & $0.7731 \pm 0.0120$ \\
54055.222 & 61\,Cyg\,B & 82.60 &  56.28 & $0.8229 \pm 0.0191$ \\
\hline
\end{tabular}
\end{table}

\subsection{Limb darkened angular diameters \label{angdiams}}

\begin{table}
\caption[]{Limb darkening coefficients of the four-parameter power law of 
Claret~(\cite{claret00}) for 61\,Cyg\,A \& B.
\label{ld-coefs}}
\begin{tabular}{lcccc}
\hline
\hline
Star & $a_1$ & $a_2$ & $a_3$ & $a_4$ \\
\hline
\noalign{\smallskip}
61\,Cyg\,A & 0.9189 & -0.5554 & 0.3920 & -0.1391 \\
61\,Cyg\,B & 1.2439 & -1.5403 & 1.3069 & -0.4420 \\
\hline
\end{tabular}
\end{table}

In order to estimate the angular diameter from the measured visibilities it is necessary to know
the intensity distribution of the light on the stellar disk, i.e. the limb darkening (LD).
We selected the four-parameter LD law of Claret~(\cite{claret00}):
\begin{equation}
{I(\mu)}/{I(1)} = 1 - \sum_{k=1}^{4}{a_k(1-\mu^{\frac{k}{2}}})
\end{equation}
The $a_k$ coefficients are tabulated by this author for a wide range
of stellar parameters ($T_{\rm eff}$, $\log g$,...) and photometric bands ($U$ to $K$).
To read Claret's tables for the PHOENIX models, the effective temperatures of 61\,Cyg\,A and B were rounded 
to $T_{\rm eff}(A) = 4400$\,K and $T_{\rm eff}(B) = 4000$\,K, with a metallicity rounded to solar ($\log[M/H]=0.0$),
a turbulent  velocity $V_T=2$\,km/s, and a surface gravity rounded to $\log g = 5.0$.
From the masses we derive in Sect.~\ref{modelfit} (listed in Table~\ref{params}) and the measured
radii (Sect.~\ref{radii}), the effective gravities are $\log g = 4.71$ and $\log g = 4.67$, respectively
for 61\,Cyg~A and B. The values of the four $a_i$ parameters corresponding to 
Claret's intensity profiles of the two stars in the $K$ band are given in Table~\ref{ld-coefs}.

The FLUOR instrument bandpass corresponds to the $K'$ filter ($1.9\leq \lambda \leq 2.3\;\mu\mathrm{m}$). 
An effect of this relatively large spectral bandwidth is that several spatial frequencies are simultaneously 
observed by the interferometer. This effect is known as bandwidth smearing.
It is usually negligible for $V^2 \ge 40$\%, but not in our case as the measured visibilities
are closer to the first minimum of the visibility function. 
To account for this effect, the model visibility is computed for regularly spaced wavenumber spectral bins 
over the $K'$ band, and then integrated to obtain the model visibility. This integral is computed numerically and 
gives the model $V^2$ as a function of the projected baseline $B$ and the angular diameter $\theta_{\rm LD}$. 
A simple $\chi^2$ minimization algorithm is then used to derive $\theta_{\rm LD}$, the 1\,$\sigma$ error bars 
and the reduced $\chi^2$. For more details about this fitting procedure, the reader is referred to Kervella et 
al.~(\cite{kervella03b}) or Aufdenberg et al.~(\cite{aufdenberg05}).

In the infrared $K$ band, the LD is much weaker than in the visible, and for instance
the difference in the LD of 61\,Cyg\,A and B is small.
It can be quantified by the ratio $\beta$ of the LD angular diameter $\theta_{\rm LD}$ and
the equivalent uniform disk angular diameter $\theta_{\rm UD}$. From Claret's models,
we obtain $\beta(A) = 1.0305$ and $\beta(B) = 1.0263$. The difference between the two
stars is 0.4\% for $\Delta T_{\rm eff} \approx 400$\,K. To account for the
uncertainty on the effective temperatures, we thus included a $\pm 0.2$\% systematic
uncertainty on the LD angular diameters of the two stars.
The LD being in any case small in the $K$ band, this appears as a reasonable assumption.

\begin{figure*}[t]
\centering
\includegraphics[width=8.6cm]{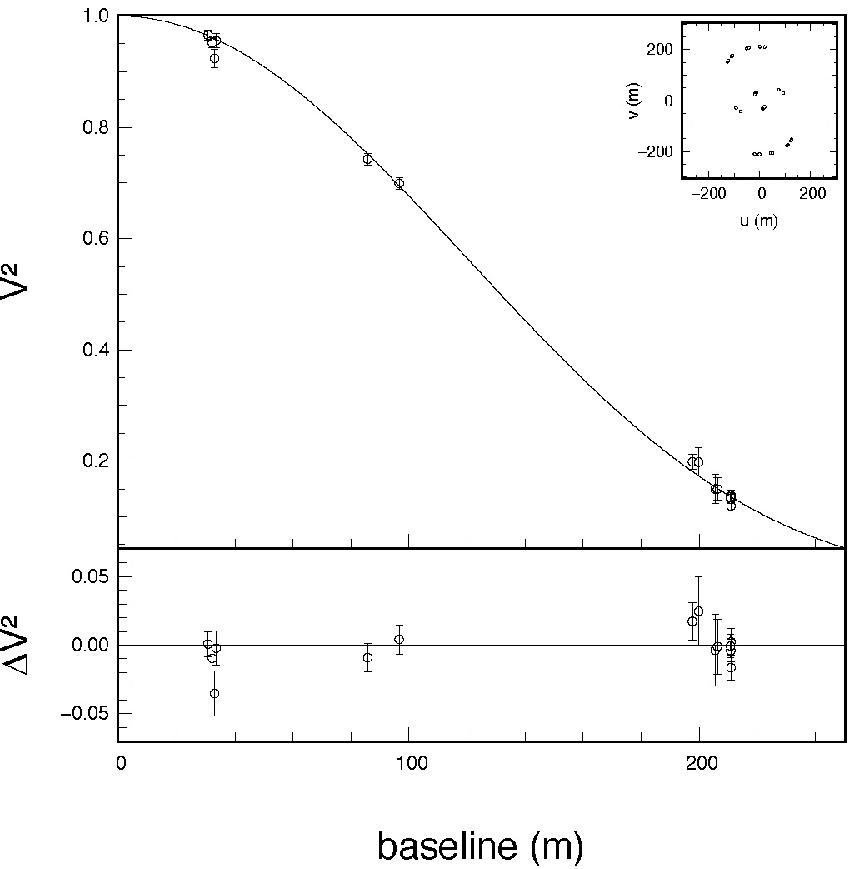} \hspace{0.5cm}
\includegraphics[width=8.6cm]{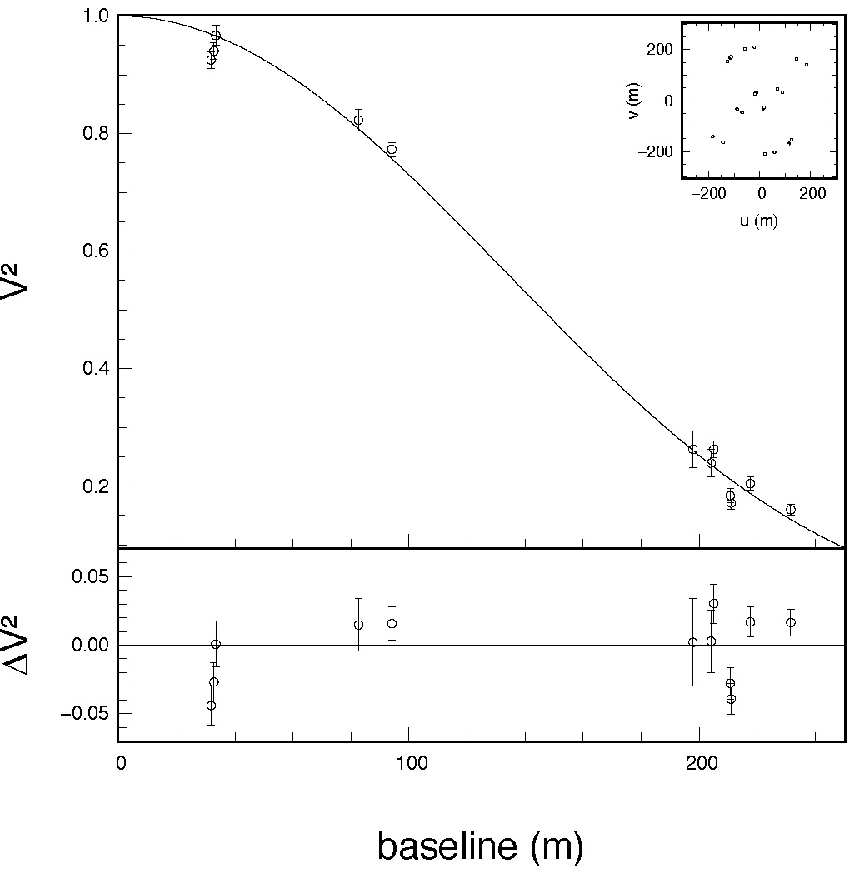}
\caption{Visibility data and adjusted limb darkened disk model visibility curve for 61\,Cyg\,A 
(left) and B (right) 
(see also Table~\ref{visib-table}).
\label{visib-curves}}
\end{figure*}

There is also an  uncertainty on the effective wavelength of the instrument.
As the measurements are differential in nature (between the scientific targets
and the calibrators), this impacts the measurement only as a second order effect.
The true effective wavelength of  FLUOR instrument was calibrated 
by observing with high accuracy the binary star $\iota$\,Peg (M\'erand et al.~\cite{merand08}, in prep.).
The relative uncertainty introduced on the angular diameters of 61\,Cyg\,A \& B is 
estimated to $\pm 0.1$\%, i.e. negligible compared to the statistical uncertainties (0.7\% and 1.4\%).
Both the LD and wavelength uncertainties were added quadratically to the errors from the fitting to arrive at the final uncertainties of the LD angular diameters.
The result of the fits are presented for 61\,Cyg\,A \& B in Figs.~\ref{visib-curves}. 
The reduced $\chi^2$ are 1.0 and 3.6, respectively for A and B.
From the CHARA/FLUOR visibilities, we derive the following LD angular diameters:

\begin{equation}
\theta_{\rm LD} ({\rm 61\,Cyg\,A}) = 1.775 \pm 0.013\ {\rm mas,}
\end{equation}
\begin{equation}
\theta_{\rm LD} ({\rm 61\,Cyg\,B}) = 1.581 \pm 0.022\ {\rm mas.}
\end{equation}

\subsection{Angular diameter discussion \label{discussion}}

We can compare the measured LD angular diameters to the expected values for these stars from 
the surface brightness-color 
relations calibrated by Kervella et al.~(\cite{kervella04b}). Using the $(B,B-L)$ relation 
(photometry from Ducati et al.~\cite{ducati02}), 
we obtain the following predicted values: $\theta_{\rm LD} ({\rm 61\,Cyg\,A}) = 1.813 \pm 0.019$\,mas, and
$\theta_{\rm LD} ({\rm 61\,Cyg\,B}) = 1.704 \pm 0.018$\,mas. 
While the agreement is good for A (within $\approx 1\,\sigma$), the predicted size for B is 
$4\,\sigma$ larger than the measured value.
Both our measured angular diameter value for 61\,Cyg\,B and this estimate from surface 
brightness are significantly different from the measurement obtained by
Lane \& Colavita~(\cite{lane03}) using the Palomar Testbed  Interferometer. 
They obtained an angular diameter of $\theta_{\rm LD} ({\rm 61\,Cyg\,B}) = 1.94 \pm 0.009\ {\rm mas}$, that 
corresponds to a difference of $+8\,\sigma$ from our CHARA/FLUOR measurement,
and $+11\,\sigma$ from  the surface brightness predicted values. 
One should note that the measurement of the angular size of the star was not the primary focus 
of their work (in particular, no spatial filter was used for this measurement,
resulting in a more difficult calibration of the visibilities).

These discrepancies combine with the larger observed dispersion of the CHARA/FLUOR 
measurements (compared to A) to indicate a possible additional contributor in the interferometric field of view around B 
($\approx$1"). Moreover, 61\,Cyg\,B shows a slight visibility deficit at our shortest baseline which corresponds 
to a $1.0 \pm 0.4$\% photometric excess. 
These observations can be explained if B is surrounded by a disk, or by the presence of a faint 
companion. Interestingly, an 8\,M$_{\rm J}$ companion was proposed by
Strand~(\cite{strand43}, \cite{strand57}), based on astrometric measurements.
This  possibility was suggested again by Deich \& Orlova~(\cite{deich77}), Deich~(\cite{deich78}),
and recently by  Gorshanov et al.~(\cite{gorshanov06}), based on astrometry of the two stars.
One should note however that a giant planet with a mass around 10\,M$_J$ and the age
of 61\,Cyg will be extremely faint, according for instance to the models by Baraffe et al.~(\cite{baraffe03}),
with an absolute $K$ band magnitude of M$_K \approx 30$, and an apparent magnitude of $m_K \approx 28$.
It thus appears unlikely that such a planet can influence the CHARA/FLUOR interferometric measurements.
Moreover, using radial velocimetry, Walker et al.~(\cite{walker95}) excluded the presence
of low mass companions around both components of 61\,Cyg down to a few Jupiter
masses, up to periods of about 30\,years.

As our present data set is too limited to investigate this possibility, we postpone this dicussion to a
forthcoming paper.
The dispersion of the $V^2$ measurements observed on B is in any case 
taken into account in the bootstrapped error bars of the LD angular diameters and 
will therefore not affect our modeling anaysis.

\subsection{Linear photospheric radii \label{radii}}

We assumed the following parallax values for the two stars:
\begin{equation}
\pi({\rm 61\,Cyg\,A}) = 286.9 \pm 1.1\ {\rm mas,}
\end{equation}
\begin{equation}
\pi({\rm 61\,Cyg\,B}) = 285.4 \pm 0.7\ {\rm mas.}
\end{equation}
They are taken respectively from van Altena et al.~\cite{vanaltena95}) and
the {\it Hipparcos} catalogue~(ESA~\cite{esa97}). As a remark, van de Kamp~(\cite{vandekamp73})
obtained slightly different values
($\pi_A = 282.8 \pm 1.3$\,mas and $\pi_B = 288.4 \pm 1.6$\,mas)
using photographic plates collected over the 1912-1972 period
(see also van de Kamp \cite{vandekamp53}). From the reprocessing
of the original {\it Hipparcos} measurements, Van Leeuwen~(\cite{vanleeuwen07a}; \cite{vanleeuwen07b})
obtained the following parallaxes:
$\pi_A = 286.8 \pm 6.8$\,mas and $\pi_B = 285.9 \pm 0.6$\,mas, compatible with our assumed values.

From the combination of the limb darkened angular diameters and trigonometric parallaxes, 
we derive the following photospheric linear radii:
\begin{equation}
{\rm R}({\rm 61\,Cyg\,A}) = 0.665 \pm 0.005\ {\rm R}_{\odot}, \nonumber
\end{equation}
\begin{equation}
{\rm R}({\rm 61\,Cyg\,B}) = 0.595 \pm 0.008\ {\rm R}_{\odot}. \nonumber
\end{equation}
The relative uncertainties on the radii are therefore $\pm 0.8$\% and $\pm 1.4$\% respectively for A and B.
Thanks to the high precision of the parallaxes (0.38\% and 0.25\%), the radius accuracy
is limited by the precision of the LD angular diameter measurements.

\section{Additional observational constraints \label{constraints}}

\subsection{Masses}

In spite of the proximity and large semi-major axis ($a=24.65\arcsec$) of 61\,Cyg,
its very long orbital period of 6-7 centuries makes
it very difficult to determine accurate dynamical masses.
Quoting Zagar~(\cite{zagar34}), Baize~(\cite{baize50}) reports a period of 692\,years, while
Cester, Ferluga \& Boehm~(\cite{cester88}) quote a period of 720\,yrs and equal masses
around $0.5 \pm 0.1\,M_\odot$ for both components.
Van de Kamp~(\cite{vandekamp54}) proposed masses of 0.58 and $0.54\,M_\odot$ for
61\,Cyg\,A and B, respectively.
From the orbit determined by Worley \& Heintz~(\cite{worley83}), Walker et al.~(\cite{walker95})
quote masses of:
\begin{equation}
M({\rm 61\,Cyg\,A}) = 0.67\,M_\odot,
\end{equation}
\begin{equation}
M({\rm 61\,Cyg\,B}) = 0.59\,M_\odot.
\end{equation}
While these values are rather uncertain, we will use them as our starting guesses
for the CESAM2k modeling. As a comparison, the masses proposed by
Gorshanov et al.~(\cite{gorshanov06}) of $M({\rm 61\,Cyg\,A}) = 0.74 \pm 0.13\,M_\odot$
and $M({\rm 61\,Cyg\,B}) = 0.46 \pm 0.07\,M_\odot$ have a larger ratio,
although the total mass of the system ($M_{A+B}=1.20 \pm 0.15\,M_\odot$) is compatible
with Walker et al.'s value.

\subsection{Effective temperature \label{teffs-sect}}

\begin{table}
\caption[]{Apparent magnitudes of 61\,Cyg\,A and B from ESA~(\cite{esa97}) and Ducati~(\cite{ducati02}).
The uncertainties that were not available were set arbitrarily to $\pm 0.05$\,mag.
\label{magnitudes}}
\begin{tabular}{lcc}
\hline\hline
Band  &  61\,Cyg\,A  &  61\,Cyg\,B \\
\hline
\noalign{\smallskip}
$U$     &   $7.48 \pm 0.06$   &  $8.63 \pm 0.05$ \\ 
$B$     &   $6.27 \pm 0.03$   &  $7.36 \pm 0.02$ \\ 
$V$     &   $5.20 \pm 0.03$   &  $6.05 \pm 0.01$ \\ 
$R$     &   $4.17 \pm 0.06$   &  $4.87 \pm 0.05$ \\ 
$I$     &   $3.52 \pm 0.06$   &  $4.05 \pm 0.05$ \\ 
$J$     &   $3.10 \pm 0.05$   &  $3.55 \pm 0.02$ \\ 
$H$     &   $2.48 \pm 0.08$   &  $2.86 \pm 0.04$ \\ 
$K$     &   $2.35 \pm 0.03$  &  $2.71 \pm 0.05$ \\ 
$L$     &   $2.28 \pm 0.06$   &  $2.61 \pm 0.05$ \\ 
\hline
\end{tabular}
\end{table}

\begin{figure}[t]
\centering
\includegraphics[width=8.6cm]{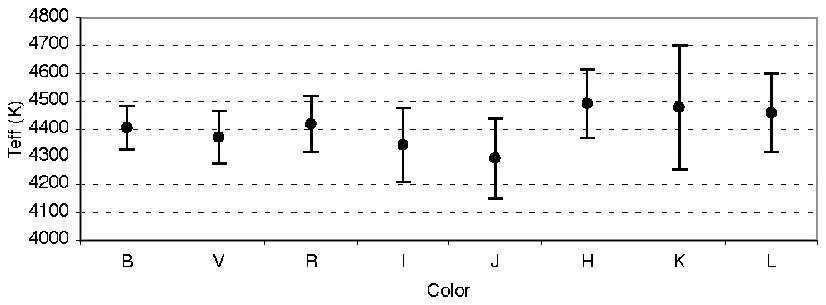}
\includegraphics[width=8.6cm]{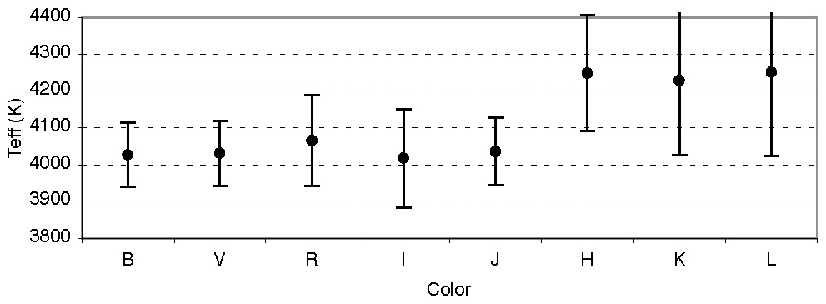}
\caption{Effective temperature estimates for 61\,Cyg\,A (top) and B (bottom) based on their apparent
magnitudes and the surface brightness-color relations by Kervella et al.~(\cite{kervella04b}).
\label{teffs}}
\end{figure}

Knowing the LD angular diameter of the stars, it is possible to invert the surface brightness-color (SBC)
relations calibrated by Kervella et al.~(\cite{kervella04b}) to retrieve
the effective temperature of a star from any of its apparent magnitudes. Applying this
method to 61\,Cyg\,A and B (Fig.~\ref{teffs}) yields consistent effective temperatures for the $BVRI$ bands of:
\begin{equation}
T_{\rm eff}({\rm 61\,Cyg\,A}) = 4400 \pm 100\,{\rm K},
\end{equation}
\begin{equation}
T_{\rm eff}({\rm 61\,Cyg\,B}) = 4040 \pm 80\,{\rm K}.
\end{equation}
The apparent magnitudes of the two stars (Table~\ref{magnitudes})
were taken from the {\it Hipparcos} catalogue (ESA~\cite{esa97}) and the catalogue compiled by
Ducati~(\cite{ducati02}). Thanks to the proximity of 61\,Cyg,
we can neglect interstellar extinction.
As a remark, the average effective temperature retrieved from the
$HKL$ bands are higher than from the visible bands, respectively by 80 and 200\,K
for 61\,Cyg\,A and B. It may be a consequence of the
activity of the stars, but more likely, it could be due to the relatively poor photometry available
in these bands. The two stars are in particular too bright to have accurate magnitudes in the 2MASS
catalogue (Skrutskie et al.~\cite{skrutskie06}). For this reason, we consider the 
$T_{\rm eff}$ estimates based on the visible photometry ($BVRI$) more reliable.
The temperatures derived from the $J$ band are however in good agreement
with these average values.
The derived temperatures are identical, within their statistical uncertainties,
to those computed from the temperature scale of
Alonso, Arribas \& Mart\'inez-Roger~(\cite{alonso96}): 4340\,K and 3980\,K.

For comparison, Luck \& Heiter~(\cite{luck05}) determined higher values of
$T_{\rm eff} = 4640$\,K and 4400\,K for 61\,Cyg\,A and B from spectroscopy.
More generally, different techniques result in significantly different $T_{\rm eff}$ values for
61\,Cyg\,A and B, and discrepancies as large as 300K are found in the literature.
This may be a consequence of the activity of the two stars (see e.g. Hall, Lockwood \& Skiff~\cite{hall07}).
61\,Cyg\,A is also classified as a BY\,Dra type variable, showing a low amplitude rotational modulation
of its light curve due to star spots and chromospheric activity (B\"ohm-Vitense~\cite{bohm07}).

\subsection{Luminosity}

The luminosity of the two stars of 61\,Cyg can be computed using two methods:
(1) combining the effective temperature and the interferometric radius,
(2) from the apparent magnitude, the bolometric correction, and the parallax.

From the effective temperatures listed in Sect.~\ref{teffs-sect} and the radii
measured with CHARA/FLUOR (Sect.~\ref{radii}), a straightforward application of
Stefan-Boltzmann's law ($L=4 \pi R^2 \sigma T_{\rm eff}^4$) gives 
$L({\rm 61\,Cyg\,A}) = 0.149 \pm 0.006\,L_\odot$, and
$L({\rm 61\,Cyg\,B}) = 0.085 \pm 0.004\,L_\odot$.

We can also use the $K$ band bolometric corrections (BC$_K$)
from Houdashelt, Bell \& Sweigart~(\cite{houdashelt00})
to retrieve the bolometric magnitude from the apparent $K$ band magnitudes of
the stars (Table~\ref{magnitudes}).
An interpolation of Houdashelt et al.'s tables gives BC$_K(A)=2.152$ and BC$_K(B)=2.433$.
Considering the uncertainty on the metallicity of the two stars, and the difference observed
between the visible and infrared effective temperature estimates (Sect.~\ref{teffs-sect}),
we attributed a $\pm 0.07$\,mag error bar to these bolometric corrections.
Together with the parallaxes, they correspond to absolute bolometric magnitudes of
$M_{\rm bol}(A) = 6.80 \pm 0.08$ and $M_{\rm bol}(B) = 7.42 \pm 0.09$,
and the following luminosities (with $M_{\rm bol}(\odot)=4.75$):
\begin{equation}
L({\rm 61\,Cyg\,A}) = 0.153 \pm 0.010\,L_\odot,
\end{equation}
\begin{equation}
L({\rm 61\,Cyg\,B}) = 0.085 \pm 0.007\,L_\odot.
\end{equation}
These figures are in excellent agreement with those computed from the effective
temperature and the radius.
In our subsequent modeling of the two stars, we will use simultaneously
as constraints the different available observables. However, various correlations
exist between these observables, for instance between $R$, $L$ and $T_{\rm eff}$
through Stefan-Boltzmann's law, and we prefer to limit the influence of these correlations
as much as possible. However, this was not possible with all parameters as,
for instance, the parallax plays a role in the derivation of most of them.
As the bolometric corrections have the advantage of being independent from the radius
measurement, we retain them for our modeling of the two stars.

\subsection{Metallicity}

An error of 0.10\,dex on the metallicity translates into an error of 160\,K on the effective temperature,
so a good knowledge of the metallicity is important. 

A number of metallicity estimates can be found in the literature.
Zboril \& Byrne~(\cite{zboril98}) determined the metallicity of 61\,Cyg\,A to
be $\left[ {\rm M/H} \right] = -0.3 \pm 0.2$, from high-resolution spectroscopy and LTE atmosphere models.
They also found a mean metallicity index $\left[ {\rm M/H} \right] = -0.2$ for a sample of 18 K and M
field stars. Tomkin \& Lambert~(\cite{tomkin99}) obtained relatively low iron abundances of
$\left[ {\rm Fe/H} \right] = -0.43 \pm 0.10$ for 61\,Cyg\,A and
$\left[ {\rm Fe/H} \right] = -0.63 \pm 0.10$ for 61\,Cyg\,B,
based on equivalent line widths and LTE model atmospheres.
Using a similar method, Affer et al.~(\cite{affer05}) measured $\left[ {\rm Fe/H} \right] = -0.37 \pm 0.19$
for 61\,Cyg\,A. 

For the present study, we choose to rely on the work by Luck \& Heiter~(\cite{luck05}),
who determined:
\begin{equation}
\left[ {\rm Fe/H} \right]({\rm 61\,Cyg\,A}) = -0.20 \pm 0.10
\end{equation}
\begin{equation}
\left[ {\rm Fe/H} \right]({\rm 61\,Cyg\,B}) = -0.27 \pm 0.19
\end{equation}
For the abundance
analysis, they used plane-parallel, line-blanketed, flux-constant, LTE, MARCS model atmospheres.
These models are a development of the programs of Gustafsson et al.~(\cite{gustafsson75}).
We selected these values, as they are well in line (within 1 to $2\,\sigma$) with the measurements
obtained by other authors. In addition, we used the MARCS models for the atmosphere description
of our evolutionary modeling (Sect.~\ref{modelfit}), so Luck \& Heiter's determination appears
as a natural choice for consistency. 

\begin{table}
\caption[]{Observational constraints (upper part),
and best-fit parameters derived from our CESAM2k modeling of 61\,Cyg\,A \& B (lower part).
See text for the corresponding references.
\label{params}}
\begin{tabular}{lcc}
\hline\hline
Parameter  &  61\,Cyg\,A  &  61\,Cyg\,B \\
\hline
\noalign{\smallskip}
$\pi$ (mas)  &   $286.9 \pm 1.1$   &  $285.4 \pm 0.7$ \\ 
$\left[ {\rm Fe/H} \right]$ & $-0.20 \pm 0.10$ &  $-0.27 \pm 0.19$ \\
R (R$_\odot$)   &   0.665 $\pm$ 0.005  &  0.595 $\pm$ 0.008 \\ 
$T_{\rm eff}$ (K)  &   $4\,400 \pm 100$   & $4\,040 \pm 80$ \\ 
$L$ ($L_\odot$)     &  $0.153 \pm 0.010$      &   $0.085 \pm 0.007$ \\ 
\hline
\noalign{\smallskip}
Initial He content Y$_{ini}$ &   0.265   &  0.265 \\
Initial $\left[ {\rm Z/H} \right]$ (dex)  &  - 0.10   &  -0.10 \\
Final $\left[ {\rm Z/H} \right]$ (dex) &  - 0.15    &  - 0.15  \\
Age (Gyr) &  6.0 $\pm$ 1.0  &   6.0 $\pm$ 1.0 \\
Mass (M$_\odot$)    &  0.690   &  0.605 \\
$\alpha$ (MLT convection) &   1.2    &  0.8   \\
\hline
\end{tabular}
\end{table}
%

\section{Modeling with CESAM2k \label{modelfit}}

The observational constraints and parameters used to construct our
CESAM2k evolutionary models (Morel~\cite{morel97}; Morel \& Lebreton~\cite{morel07})
are summarized in Table~\ref{params}.

As a binary system, components A and B of 61\,Cyg must have the same age, initial helium content,
and metallicity (assuming that the system formed as a binary).
By comparison with our previous modeling efforts (see e.g. Th\'evenin et al.~\cite{thevenin05}),
we implemented a few modifications for the present work. The equation of state now includes Coulomb
corrections due to low masses of these stars and the corresponding high density.
We used MARCS models (Gustafsson et al.~\cite{gustafsson03,gustafsson07,gustafsson08})\footnote{{\it http://www.marcs.astro.uu.se/}} for the atmosphere description. The treatment of convection
makes use of a new modified routine ({\tt conv\_a0}) based on the classical mixing length theory
(hereafter MLT). The mixing length vanishes at the limit between convective and radiative
zones\footnote{see the documentation of CESAM2k on the Helas website
{\it http://helas.group.shef.ac.uk/solar\_models.php}}.
Note that our small values of the $\alpha$ parameter (that describes the convection in MLT) are in good
agreement with the systematic trend of this parameter in low mass stars, where its value decreases steeply with mass. 
For instance, if for the Sun the most common quoted value of $\alpha$ is about 1.9 (Morel et al.~\cite{morel99}), 
Yildiz et al.~(\cite{yildiz06}) obtain for low mass stars values such as 1.2 for 0.8\,M$_{\odot}$ or 
1.7 for 0.9\,M$_{\odot}$. Compared to these values, our values of $\alpha$ of 0.8 for 61\,Cyg\,B ($0.6\,M_{\odot}$)
and 1.2 for 61\,Cyg\,A ($0.7\,M_{\odot}$) appear reasonable.

We define the stellar radius of a model as the bolometric one, which is equivalent to the interferometric
definition of the limb darkened angular diameter.
The microscopic diffusion of chemical species is taken into account according to Burgers~(\cite{burgers69}),
using the resistance coefficients of Paquette et al.~(\cite{paquette86}).
According to the prescription of Morel \& Th\'evenin~(\cite{morel02}), we introduce an additional mixing 
parameterized by Re$\nu$ $\approx1$. 
This parameterization is not important for the stars 61\,Cyg~A \& B because the diffusion is 
inefficient for low mass stars with solar metallicity.

The adopted metallicity [Z/X], which is an input parameter for the evolutionary computations, is 
given by the iron abundance measured in the atmosphere with the help of the following 
approximation: 
$\log ({\rm Z/X}) \approx [{\rm Fe/H}] + \log ({\rm Z/X})_\odot$ . 
We use the solar mixture of Grevesse \& Noels~(\cite{grevesse93}): $(Z/X)_\odot = 0.0245$.
The evolutionary tracks are initialized at the Pre-Main Sequence stage ; therefore, the ages are 
counted from the ZAMS.

\begin{figure}[t]
\centering
\includegraphics[width=7.cm, height=8.7cm, angle=-90]{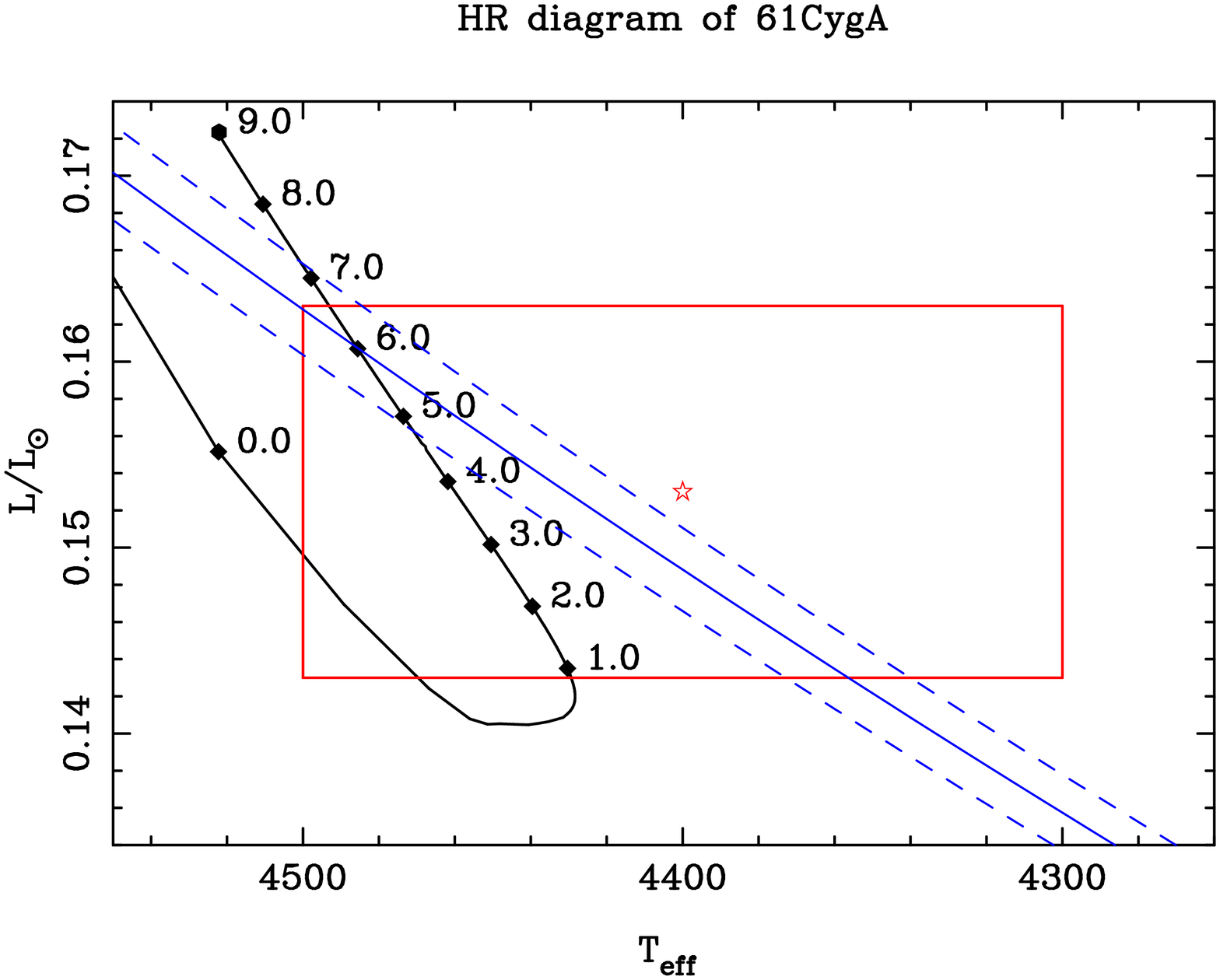}
\includegraphics[width=7.cm, height=8.7cm, angle=-90]{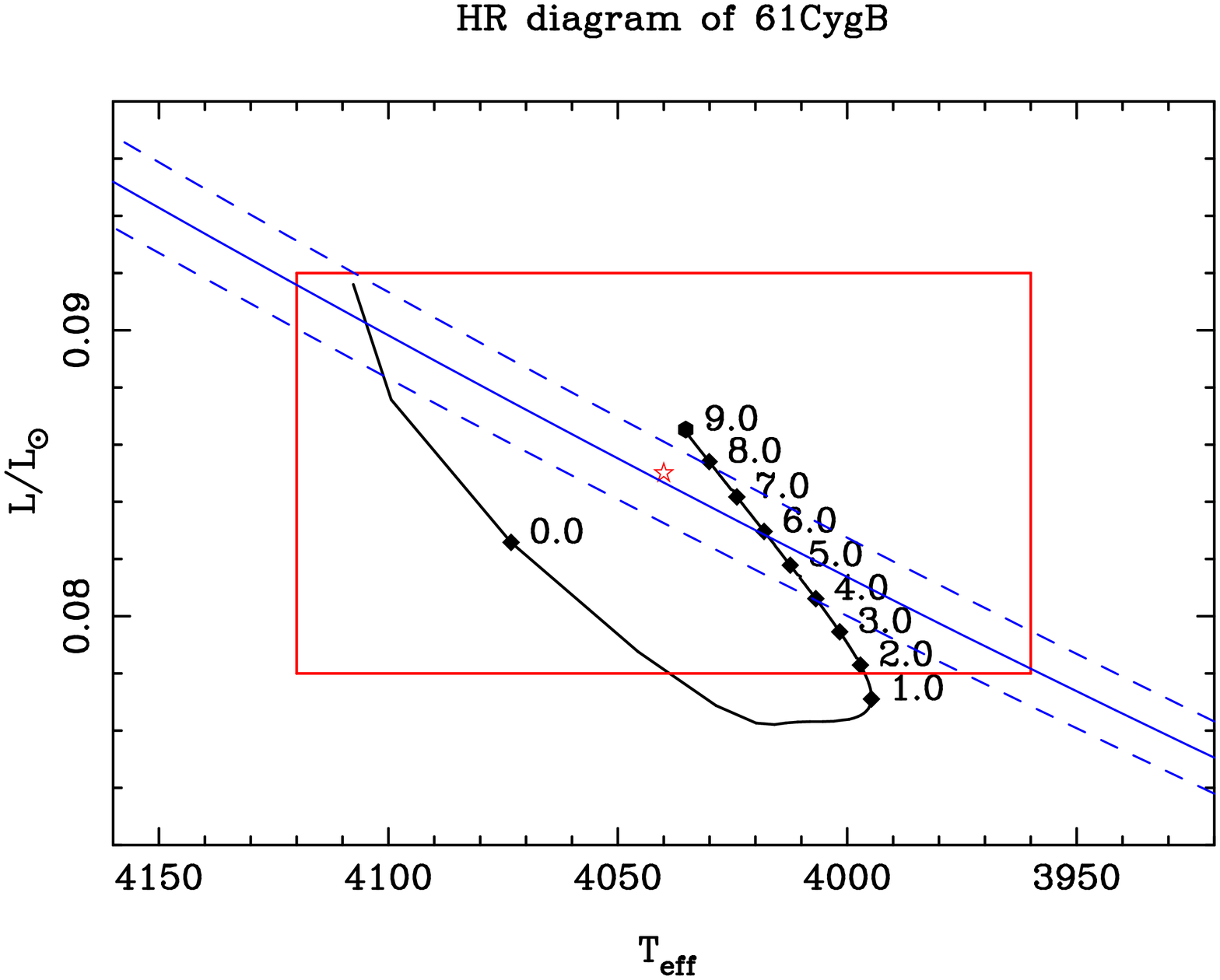}
\caption{Evolutionary tracks in the H-R diagram for 61\,Cyg\,A (top) and B (bottom). The labels indicate the age in Gyr relative to the ZAMS. The rectangular box represents the classical $L-T_{\rm eff}$ error box, and the diagonal lines represent the radius and its uncertainty. \label{HR_61Cyg} }
\end{figure}

To fit the observational constraints ($T_{\rm eff}$, $L$ and surface metallicity [Z/X]$_{surf}$)
with corresponding results of various computations, we adjust the main stellar
modeling parameters: mass, age and metallicity.
In Fig.~\ref{HR_61Cyg}, the rectangular error boxes correspond to
the values and accuracies of the $T_{\rm eff}$ and $L$ parameters quoted in Table~\ref{params}.
The values of the radii reported in the present work select very narrow diagonal sub-areas in these
error boxes. The new measurements of radii are thus particularly discriminating for the models,
as previously noticed by Th\'evenin et al.~(\cite{thevenin05}) and Creevey et al.~(\cite{creevey07}).

We adopt an initial helium content of Y$_{ini}$=0.265 which correspond to a
slightly metal poor disk star. We then tried to fit the evolutionary models
within the error boxes of both stars with the masses proposed
by Gorshanov et al.~(\cite{gorshanov06}) of 0.74 and 0.46~M$_\odot$.
In this process, we explored a range of abundance Z/X by changing the observed value by $\pm 0.15$\,dex
and Y by $\pm 0.01$\,dex. We also tried varying the masses by $\pm 0.05\,M_\odot$.
No combination of the three parameters produced evolutionary models reaching the
error boxes of the HR diagram, that are strongly constrained by the radius.
The variation of the $\alpha$ parameter of the convection depth did not help either.
From this we conclude that the masses proposed by Gorshanov et al.~(\cite{gorshanov06})
are not reproducible by our modeling within $\pm 0.05\,M_\odot$. For this reason,
we decided to use as a starting value the older mass determination by Walker et al.~(\cite{walker95})
of 0.67 and 0.59\,M$_\odot$. The convergence of our models towards the observations is significantly
better and we could refine these masses to the values listed in Table~\ref{params}.

We selected as the most plausible models those satisfying
first the luminosity and radius constraints and second the effective temperature constraint.
The corresponding parameters are given in Table~\ref{params}. 
The models of 61\,Cyg\,A and B converge simultaneously to the radii-limited uncertainty boxes for
an age of $6.0 \pm 1.0$\,Gyr. This is significantly older than the 2.1--1.9\,Gyr estimate of
Barnes~(\cite{barnes07}), that was derived from the measured rotation period
of the stars (gyrochronology). The chromospheric age quoted by this author of 2.4--3.8\,Gyr also
appears lower than our value.

\section{Asteroseismic frequency predictions \label{astero}}

Late-type bright binaries are rare but represent, once analysed with asteroseismic constraints, 
an excellent challenge for internal structure models to derive the age and helium content in 
the solar neighborhood.
With the improved determinations of the fundamental parameters of the 61\,Cyg system, we 
propose predictions of pulsation frequency separations useful for asteroseismic diagnostics.
To date, no detection of pulsation has been observed in these two stars. Our purpose here is not to 
predict which modes are excited but rather to compute a broad range of eigenfrequencies $\nu_{n,\ell}$ with 
radial orders $n>10$ and degrees $\ell=0,1,2,3$ up to the cut-off frequency, so that they will 
cover possible future detections and eventually help to constrain future models of these stars. 
For a thourough review of the asteroseismic concepts discussed in the present Section,
the interested reader is referred to the excellent review by Cunha et al.~(\cite{cunha}).

\begin{figure}[t]
\centering
\includegraphics[angle=90,width=9cm]{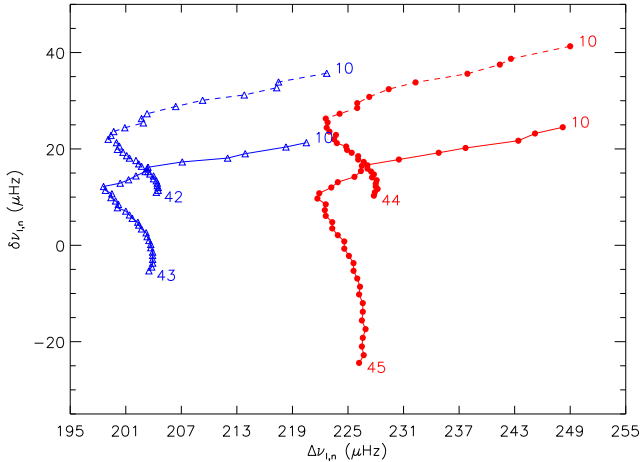}
\caption{Diagram representing the small separations $\delta\nu_{n,\ell}$ (vertical axis)
as a function of the large ones $\Delta\nu_{n,\ell}$ (horizontal axis) for the
two components 61\,Cyg\,A ($\triangle$) and B ($\bullet$).  We consider here only 
$(\delta\nu_{n,0},\Delta\nu_{n,0})$ (solid line) and
$(\delta\nu_{n,1},\Delta\nu_{n,1})$ (dashed line). The successive radial nodes $n>10$ are 
indicated on the curves. \label{separation}}
\end{figure}

The frequencies $\nu_{n,\ell}$ are calculated with a standard adiabatic code for stellar 
pulsations (e.g. Unno et al.~\cite{unno}). 
Instead of looking at the absolute values of the frequencies, we consider the so-called small 
$\delta \nu_{n\ell}= \nu_{n,\ell} - \nu_{n-1,\ell+2}$ and large separations $\Delta \nu_{n,\ell} = 
\nu_{n,\ell} - \nu_{n-1,\ell}$ which are the relevant quantities to constrain stellar properties
(e.g. Christensen-Dalsgaard~\cite{christ}). The former gives information about the stellar 
evolutionary status of the star since low degree modes have their inner turning
points close to or in the  core of the star. It is particularly
dependent on the gradient of the sound speed in the core which changes as the star evolves (e.g. 
Gough~\cite{gough86}). The large frequency separation $\Delta \nu_0$ roughly equals:
\begin{equation}
\Delta \nu_0 = \left( 2\,\int_0^R \frac{dr}{c_s} \right)^{-1}
\end{equation}
with $c_s$ the sound speed. This quantity is a measure of the propagation time inside the star.
It is proportional to the mean density which is then a strong constrain on the 
mass once the radius is determined by
interferometry (Cunha et al.~\cite{cunha}). We plot these separations in Fig.~\ref{separation} 
in a $(\delta \nu_{n,\ell},\Delta
\nu_{n,\ell} )$ diagram for the two components 61\,Cyg~A \& B.
The small and large  separations show the same behaviors with the
radial nodes for the two stars. The shifts between $\Delta\nu_{n,\ell}$ for the two stars reflect 
the difference of their mean
densities. It is interesting to note the change of sign of $\delta \nu_{n,\ell}$ for large values of 
$n$.

\section{Conclusion \label{conclusion}}

We presented high accuracy interferometric measurements of the angular diameters of the two 
nearby stars 61\,Cyg~A \& B: $\theta_{\rm LD}(A) = 1.775 \pm 0.013$\,mas 
and $\theta_{\rm LD}(B) = 1.581 \pm 0.022$\,mas, corresponding to photospheric
radii of ${\rm R}(A) = 0.665 \pm 0.005\ {\rm R}_{\odot}$ and ${\rm R}(B) = 0.595 \pm 0.008\ {\rm R}_{\odot}$.
We computed CESAM2k models that reproduce these radii as well as the other observed
properties of the two stars. 61\,Cyg~A \& B appear as very promising targets for future asteroseismic studies,
and we also derived asteroseismic frequencies potentially present in these two stars.
The detection of oscillations would bring important constraints to stellar structure
models in the cool, low-mass part of the HR diagram, where convection plays a central role.
However, in the absence of measured asteroseismic frequencies,
it appears difficult to go beyond the present modeling of the binary system.
The main reason is that the mass is
not constrained sufficiently well by the long period astrometric orbit.
We encourage asteroseismic groups to include these two stars in their observing programmes,
as the measurement of seismic parameters (in particular the mean large frequency spacing)
will bring a decisive constraint to the mass of these stars. We also found that
the value of the mixing length parameters for both stars is very well
constrained by the radius once the other parameters of the model are fixed,
in particular the mass. We reach the same conclusions as Yildiz et al.~(\cite{yildiz06})
that the $\alpha$ parameter is mass dependent at least for solar
abundance stars. The future {\it Gaia} mission (Perryman~\cite{perryman05})
will open the way to a complete calibration of several thousand binaries.
Such a large sample will give us a completely renewed view of convection
in stars (see e.g. Lebreton~\cite{lebreton05}).

\begin{acknowledgements}
The authors would like to thank all the CHARA Array and Mount Wilson Observatory day-
time and night-time staff for their support. 
The CHARA Array was constructed with funding from Georgia State University, the National 
Science Foundation, the W. M. Keck Foundation, 
and the David and Lucile Packard Foundation. The CHARA Array is operated by Georgia 
State University with support from the College of 
Arts and Sciences, from the Research Program Enhancement Fund administered by the Vice 
President for Research, and from the National Science 
Foundation under NSF Grant AST~0606958.
This work also received the support of PHASE, the high angular resolution partnership between ONERA, Observatoire de Paris, CNRS and University Denis Diderot Paris 7.
This research took advantage of the SIMBAD and VIZIER databases at the CDS, Strasbourg 
(France), and NASA's Astrophysics Data System Bibliographic Services.
\end{acknowledgements}

{}

\end{document}